\documentclass[aps,prl,preprint,showpacs,preprintnumbers]{revtex4}
\usepackage{graphics}
\def \beq{\begin{equation}}
\def \eeq{\end{equation}}
\def \beqa{\begin{eqnarray}}
\def \eeqa{\end{eqnarray}}
\def \ie{{\sl i.e.\/}}
\def \alphas{\alpha_{\scriptscriptstyle S}}
\def \bose{n_{\scriptscriptstyle B}}
\def \chiv{\chi_{\scriptscriptstyle V}}
\def \rhov{\rho_{\scriptscriptstyle EM}}
\def \prop{G_{\scriptscriptstyle EM}}
\def \fprop{{\cal G}_{\scriptscriptstyle EM}}
\def \vecp{G_{\scriptscriptstyle V}}
\def \vecpf{{\cal G}_{\scriptscriptstyle V}}
\def \vertex{C_{\scriptscriptstyle EM}}
\begin{document}

\title{The electrical conductivity and soft photon emissivity of the QCD plasma}
\author{Sourendu \surname{Gupta}}
\email{sgupta@tifr.res.in}
\affiliation{Department of Theoretical Physics, Tata Institute of Fundamental
         Research,\\ Homi Bhabha Road, Mumbai 400005, India.}

\begin{abstract}
The electrical conductivity in the hot phase of the QCD plasma is
extracted from a quenched lattice measurement of the Euclidean time
vector correlator for $1.5\le T/T_c\le3$. The spectral density in the
vicinity of the origin is examined using a method specially adapted to
this region, and a peak at small energies is seen. The continuum limit
of the electrical conductivity, and the closely related soft photon
emissivity of the QCD plasma, are then extracted from a fit to the
Fourier transform of the temporal vector correlator.
\end{abstract}
\pacs{11.15.Ha, 12.38.Gc, 12.38.Mh \hfill TIFR/TH/03-01}
\preprint{TIFR/TH/03-01, hep-lat/0301006}
\maketitle

The soft photon production rate from the plasma phase of hadronic
matter is of importance to searches for the QCD phase transition
\cite{wa98}. Consequently, there has been a long history of attempts
at perturbative computations of this rate \cite{history}. The first
lattice prediction of dilepton (off-shell photon) rates was performed
a while back \cite{dilepton}. Recently the leading order computation of
the photon production rate was completed \cite{amy}. The Kubo formula
relates the soft limit of this rate to the DC
electrical conductivity of the QCD plasma, $\sigma$.  To leading
log accuracy in the gauge coupling, $g=\sqrt{4\pi\alphas}$, one has
$\sigma\propto \alpha T/g^4\log g^{-1}$, where $\alpha$ is the fine
structure constant. The proportionality constant has been computed
recently in the leading-log approximation \cite{amy2}.  Here we report the first computation of $\sigma$
and the soft photon emissivity from a quenched lattice computation in a
region of temperature where $g$ is large and the weak-coupling approach
fails \cite{amynote}.
Our methods can also be applied to other transport problems.

The photon emissivity at temperature $T$ is related to the imaginary part
of the retarded photon propagator, {\sl i.e.\/}, the spectral density,
$\rhov^{\mu\nu}$, for the electromagnetic current correlator, through the relation
\beq
   \omega\frac{d\Omega}{d^3{\mathbf p}} = \frac1{8\pi^3}
      \bose(\omega,T) {\rhov}_\mu^\mu(\omega,{\mathbf p},T).
\label{rate}\eeq
In this work we shall take $\omega={\mathbf p}=0$, and hence obtain
the soft photon production rate. Since the EM Ward identity gives
$\rho^{00}(\omega,{\mathbf 0},T)=0$, this soft limit is related to
transport properties of the QCD plasma through the Kubo formula,
\beq
   \sigma(T) = \frac16\left.\frac{\partial}{\partial\omega}
      {\rhov}_i^i(\omega,{\mathbf 0},T)\right|_{\omega=0},
\label{cond}\eeq
where the sum is over spatial polarisations.
A lattice determination of this rate proceeds from the spectral
representation for Euclidean current correlators---
\beq
   \prop(t,{\mathbf p},T) = \int_0^\infty \frac{d\omega}{2\pi}
       K(\omega,t,T) \rhov(\omega,{\mathbf p},T),
\label{spec}\eeq
where the integral kernel $K=\exp(\omega t)\bose(\omega,T)+ \exp(-\omega
t)[1 +\bose(\omega,T)]$. $\prop$ is the product of the vector correlator
summed over all polarisations, $\vecp$, and the EM vertex factor
$\vertex=4\pi\alpha\sum_fe_f^2$, where $e_f$ is the charge of a quark
of flavour $f$. On discretising the integral it becomes clear that the
extraction of $\rhov$ from the lattice computation of $\prop$ is akin to
a linear least squares problem. The complication is that the (potentially
infinite) number of parameters to be fitted exceeds the number of data
points (which is half the number of lattice sites in the time direction,
$N_t$). The solution is to constrain the function $\rhov$ through an
informed guess \cite{baym}, and use a Bayesian method to extract it
\cite{nullsp}.
The Maximum Entropy Method (MEM) \cite{gubernatis,asakawa} along with
a free-field theory model of the spectral function has been used in
the past \cite{dilepton}.  The hard dilepton rate for $\omega/T\ge4$
is fully under control, with lattice and perturbation theory in good
agreement \cite{dilepton}. For that reason we concentrate here on the
electrical conductivity and the soft photon rate.

Correlators were investigated at $T=1.5T_c$, $2T_c$ and $3T_c$ in
quenched QCD.  The temperature range is realistic for heavy-ion
collisions. However, $g>1$ in this entire range of temperature
\cite{precise} and is therefore ineffective in the separation of length
scales upon which weak-coupling approaches depend.  In order to make
continuum extrapolations, the computations were performed on a sequence of
lattice spacings, $a=1/8T$, $1/10T$, $1/12T$ and $1/14T$, ({\sl i.e.\/},
$N_t=8$, 10, 12 and 14).  Quark mass effects were controlled by working
with staggered quarks of masses $m/T_c=0.03$ and $0.1$.  Details of the
runs, statistics, and the generation of configurations for $N_t<14$ are
described in \cite{gavai}.  For these lattice spacings the computations
were performed on two different spatial volumes in order to control
finite volume effects. For $N_t=14$ we have added runs on $14\times30^3$
lattices for $T=1.5T_c$ and $2T_c$, and on $14\times44^3$ lattices for
$T=3T_c$, generating 50 configurations separated by 500 sweeps each.
We have
measured vector correlators with two degenerate flavours of quarks.
It has been demonstrated recently that in this limit the charged and
uncharged vector correlators are identical \cite{nf11}.

Small but statistically significant differences between the lattice
results and ideal quark gas predictions for $\prop$ are observed at all
temperatures, lattice spacings, quark masses and volumes investigated. In
any lattice computation, we expect the high frequency part of $\rhov$ to
contain lattice artifacts. Moreover, physics at momenta of order $1/a$
is perturbative \cite{short} and not of interest in the
present context. We remove this physics by taking
the difference between the Euclidean temporal propagators in QCD and an
ideal quark gas (free field theory) on the same lattice---
\beq
   \Delta \prop(\omega)=\prop^{\scriptscriptstyle QCD}(\omega)-
       \prop^{ideal}(\omega),
\label{deltag}\eeq
A subtraction is needed to remove the $\omega^2$ divergent
pieces from the dispersion relations \cite{hilg}.
Other benefits accruing from this are discussed later.

We have estimated the spectral density by two classes of methods. The
first class of general techniques consist of discretising the integral
in eq.\ (\ref{spec}) into $N_\omega$ energy bins and rewriting the
equation in the form $\prop=K\rhov$ where $K$ is now an $N_t\times
N_\omega$ matrix, $\prop$ the data vector of length $N_t$ and $\rhov$
represents a vector of length $N_\omega$ \cite{integ}. For $N_\omega>N_t$
the solution is non-unique.  Additional constraints, called priors,
must then be imposed to determine them \cite{ill}.  The extraction of
the spectral density is performed in the context of Bayesian parameter
extraction. Given the data on $G$, the probability distribution function
for $\rho$ can be written using Bayes' formula
\beq
   P(\rho|G) =  P(G|\rho) P(\rho) / P(G),
\label{bayes}\eeq
where $P(A|B)$ denotes the conditional probability of $A$ given
$B$. The probability distribution function $P(\rho)$ contains the prior
information on $\rho$ that is needed for the analysis. Writing
$P(\rho|G)=\exp[-F(\rho)]$, the maximum likelihood analysis of the
probability reduces to the problem of minimizing $F$. Since $P(G)$ is
independent of $\rho$, this problem can be formulated as a minimisation
of the function
\beq
   F(\rho) = (G-K\rho)^T\Sigma^{-1}(G-K\rho) + \beta U(\rho),
\label{minim}\eeq
where the first term is the logarithm of $P(G|\rho)$ and
$P(\rho)=\exp[-\beta U]$.  The superscript $T$ denotes a transpose,
$\Sigma$ is the covariance matrix of the data, and $\beta$ is a
non-negative parameter whose choice is specified later.

The MEM technique consists of choosing some vector $\rhov^0$ and
defining $U(\rhov)=\sum_i\rhov^i[\log(\rhov^i/\rhov^{0\,i})-1]$, where the
sum is over components of the vectors. In previous works the prior
$\rhov^0$ has been chosen to be the vector spectral function in an
ideal quark gas \cite{dilepton}. Another whole class of techniques
is obtained by choosing $U(\rhov)=[{\cal L}(\rhov-\rhov^0)]^2$
where ${\cal L}$ is a non-singular matrix. The choices ${\cal L}=1$, $D$
and $D^2$ (where $D$ is a discretisation of the derivative) have
been suggested in the literature. ${\cal L}=1$ is the model that
$\Delta\rhov\equiv\rhov-\rhov^0=0$ except as forced by the data,
${\cal L}=D$ makes the {\sl a priori} choice that $\Delta\rhov$ is
constant and ${\cal L}=D^2$ is the prior choice of smooth $\Delta\rhov$
\cite{honerkamp}.

\begin{figure}[hbt]\begin{center}
   \scalebox{1.0}{\includegraphics{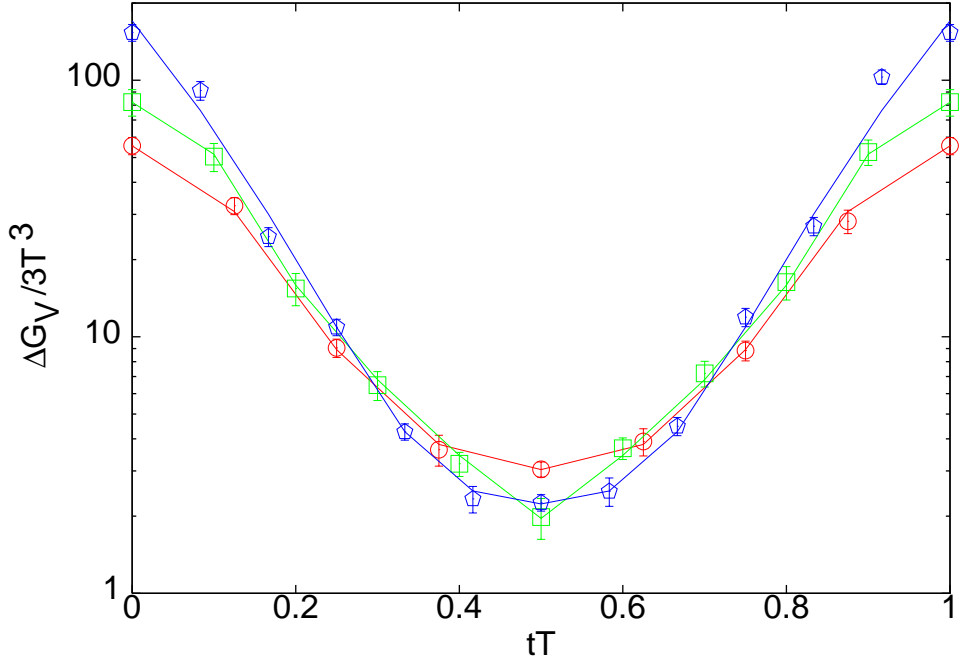}}
   \end{center}
   \caption{Bayesian fits to $\Delta\vecp(t)$ at $T=2T_c$ for $m/T_c=
      0.03$ and $N_t=8$ (circles) 10 (squares) and 12 (pentagons). The
      fits were made with $N_\omega=16$ and $0\le\omega\le4\pi T$, choosing
      ${\cal L}=1$. Changes in the fits due to variations in these algorithmic
      quantities are indistinguishable on the scale of this figure.}
\label{fg.bayesfit}\end{figure}

Such regulators have the added advantage that minimisation of the function
$F$ in eq.\ (\ref{minim}) yields the linear problem---
\beq
   \left[K^T\Sigma^{-1} K + \beta {\cal L}^T{\cal L}\right]\rhov 
           = K^T\Sigma^{-1} \prop +\beta{\cal L}^T{\cal L}\rhov^0.
\label{mineq}\eeq
Since ${\cal L}^T{\cal L}$ is positive definite, it is clear that the
term in $\beta$ on the left hand side regulates the problem, by adding
a term to $K^T\Sigma^{-1} K$ which makes the sum invertible.  Since, for
a well-determined parameter fitting problem, the value of $F$ is the
$\chi^2$ value, we choose a value of $\beta$ at which $F=N_t$ at the
minimum of $F(\beta)$, {\sl i.e.\/}, at the maximum {\sl a posteriori\/}
probability.

Considering the Bayesian problem as a field theory for the function
$\rhov$, the method of maximising the {\sl a posteriori} probability
is equivalent to a semi-classical solution. The advantage of choosing
a linear regulator is two fold. First, the search for the minimum is
simply the solution of a system of linear equations; in non-linear
minimisation it is no simple matter to correctly identify the global
minimum \cite{press}. Second, the linear problem is guaranteed to have a
single minimum, whereas a general non-linear regulator may have multiple
local minima, leading to complications analogous to the physics of
phase transitions.

\begin{figure}[hbt]\begin{center}
   \scalebox{1.0}{\includegraphics{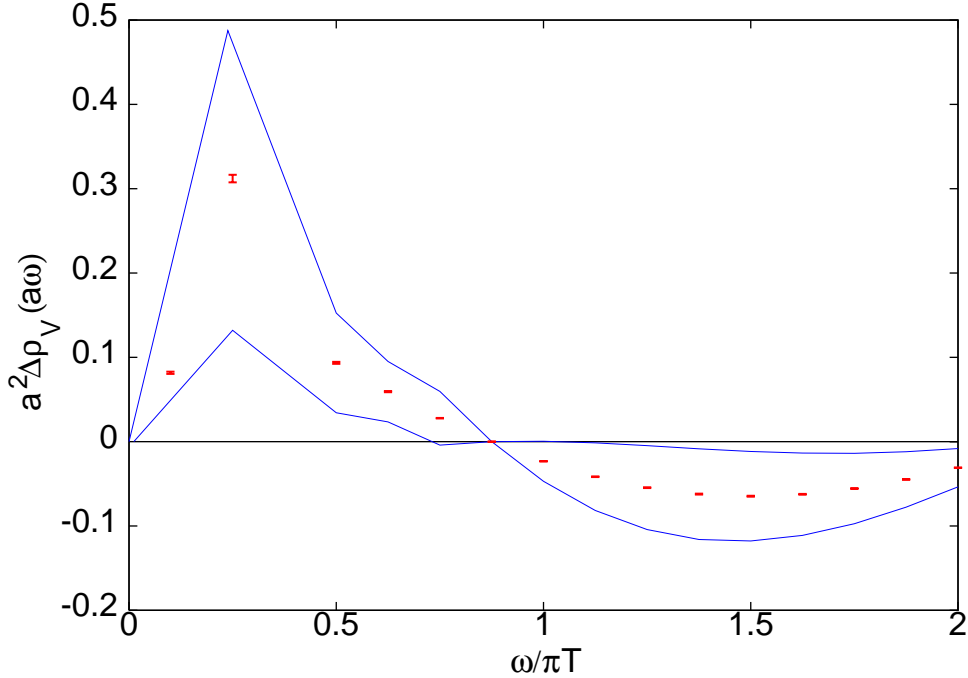}}
   \end{center}
   \caption{$\Delta\rho_{\scriptscriptstyle V}(\omega)$ obtained from
      fits to $\Delta\vecp(t)$ at $T=2T_c$ determined on a
      $12\times26^2\times48$ lattice with $m/T_c=0.03$.  Statistical
      errors obtained with $N_\omega=32$ are denoted by the bars, while
      the lines span thrice the range allowed by various systematic
      uncertainties as discussed in the text.}
\label{fg.rhov}\end{figure}

Since previous work has demonstrated that for $\omega\gg T$ lattice
computations match perturbation theory \cite{dilepton}, we focus our
attention on the region $\omega\le\pi T$.  The linear relation between
$\prop$ and $\rhov$ means that we can assume $\rhov=\rhov^0+\Delta\rhov$,
where $\rhov^0$ is the usual MEM prior in an ideal  quark gas. At small
$\omega$ this goes to zero faster than linearly in $\omega$ and hence
does not contribute to $\sigma/T$ \cite{deltafoot}.  By choosing to
work with $\Delta\prop$, this $\rhov^0$ is removed from the problem,
and we are freed to concentrate on the piece $\Delta\rhov$, which
contains all the information needed to extract $\sigma$. Then in eq.\
(\ref{mineq}) we use ${\cal L}=1$, replace $\prop$ by $\Delta\prop$
and $\rhov$ by $\Delta\rhov$, and remove the term in $\rhov^0$. The
upper limit of the integral was truncated to $\omega=2n\pi T$ and the
range divided into an uniform mesh of $N_\omega$ points. Varying $n$
and $N_\omega$ independently in the range $2\le n\le4$ and $16\le
N_\omega\le64$ has no effect on the quality of the fit to the data
(see Figure \ref{fg.bayesfit}).

Statistical errors on $\Delta\rhov$ are assigned by a bootstrap over
the measured values of $\Delta\prop$.  These are minor compared to
uncertainties arising from algorithmic parameters. The latter are
estimated by changing the integration method which is used to discretise
eq.\ (\ref{spec}), the bin size in the integration, the integration
limit, $N_\omega$, and the Bayesian prior specified by the matrix
${\cal L}$. Thrice the combined uncertainty due to these four factors
is shown as the band in Figure \ref{fg.rhov} within which $\Delta\rhov$
lies. Even with this generous allowance for uncertainties there seems to
be a peak in the spectral function at small $\omega$. For $\omega\ge\pi T$
the spectral function is roughly consistent with free field theory, but
there is some evidence of a further peak at $\omega\approx 16\pi T$. The
most important systematic uncertainty turns out to be related to control
over the limit $a\to0$. We found that the position of the peak and the
slope at the origin change in going from $N_t=8$ to 14. This phenomenon
has been noticed earlier in the context of MEM \cite{limit}. A method
which allows for better control of the continuum limit is required.

For this we utilize a second class of Bayesian methods, in which the prior
is a model of the observed bump in the soft part of the spectrum. Since
$\rhov$ is real and non-singular for real $\omega$, odd in $\omega$,
and non-negative for $\omega>0$, one can choose to work with the most
general form which gives rise to a non-vanishing electrical conductivity,
\beq
   \frac1{T^2}\Delta\rhov(z) = \frac{z\sum_{n=0}^N\gamma_nz^{2n}}{1+
      \sum_{m=1}^M\delta_mz^{2m}},
\label{model}\eeq
where $z=\omega/T$ and with all $\gamma_n$ and $\delta_m$ real and
non-negative \cite{aarts}.  The constraint that $\Delta\rhov\to0$ at
large $\omega$ is imposed by choosing $M>N$ \cite{param}. We shall use
the notation $(N,M)$ to denote a particular choice of $N$ and $M$. Bayesian
techniques for parameter estimation then proceed by choosing {\sl a
priori} probability distributions for each parameter \cite{lepage}.

The parameters in eq.\ (\ref{model}) are most conveniently extracted by
fitting to the Fourier transform
of eq.\ (\ref{spec}) over the Euclidean time $t$ \cite{dolan}
\beq
   \fprop(\omega_n,{\mathbf p},T) = \oint\frac{d\omega}{2i\pi}
         \frac{\rhov(\omega,{\mathbf p},T)}{\omega-\omega_n},
\label{eucl}\eeq
where the Euclidean frequencies are $\omega_n=2in\pi T$, ($1\le n\le
N_t$ on a lattice) and the path of integration over complex $\omega$
runs over the real line and is closed in the upper half-plane.  The
form of $\rhov$ in eq.\ (\ref{model}) can then be used to express the
Fourier coefficients in terms of the parameters, which can be
determined either through a least squares method if $1+N+M<N_t/2$, or a
Bayesian method when there are more parameters than data.

\begin{figure}[htb]\begin{center}
   \scalebox{1.0}{\includegraphics{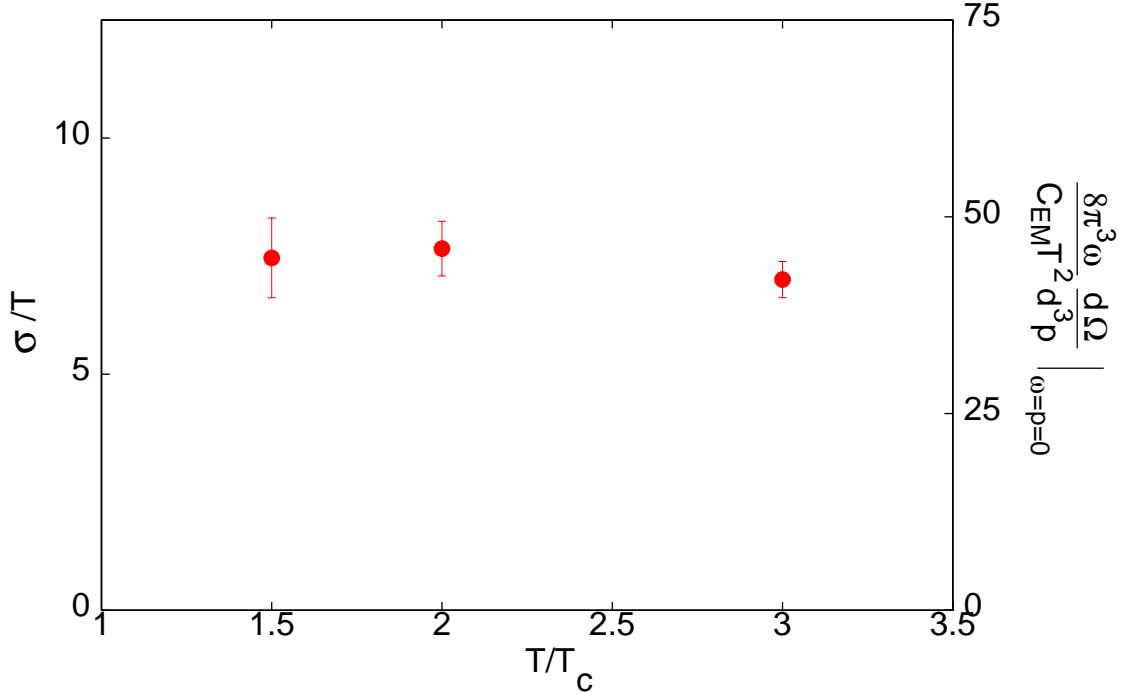}}
   \end{center}
   \caption{The electrical conductivity of the QCD plasma, $\sigma$, as
      a function of the temperature, $T$. The dimensionless quantity
      related to the soft photon emission rate shown on the right hand
      y-axis equals $6\sigma/T$ \protect\cite{deltafoot}. The bars denote
      statistical errors in the fit to the form in eq.\
      (\protect\ref{model}) with $M=2$ and $N=1$.}
\label{fg.phot}\end{figure}

A particular simplification occurs for $(0,M)$, since
$\gamma_0$ is the only parameter that contributes for $\omega_n=0$. In
all these cases $\sigma/T=\vertex(\chiv-\chiv^0)/3T^2$, where
$\chiv=\vecpf(0,{\mathbf 0},T)$ is the vector meson susceptibility
\cite{gupta} obtained in QCD and $\chiv^0$ is the same quantity for an
ideal quark gas on the same lattice \cite{arvind}.  Since the remaining
parameters do not appear in this expression, their prior probabilities
can be integrated out of the problem, without any assumptions about
them. Such a marginalisation of the prior distribution is a general
technique which has been demonstrated on other problems in the past
\cite{for}.

The extraction using $N=0$ must be insufficient for $T>10^{11}T_c$,
since it does not reproduce the parametric dependence of $\sigma$ on $g$
in weak-coupling theory, which is expected to work at these temperatures.
This can be improved by allowing for other values of $(N,M)$.  We
have investigated the stability of our results by going to $(1,2)$.
Such a multiparameter fit moves the result up by 7\%, which is
within the statistical uncertainty. The electrical conductivity is thus
reasonably stable, although it would be interesting in future to
investigate its stability further, especially by using larger values of
$N_t$.

In principle, such an extraction of parameters other than $\gamma_0$ in
eq.\ (\ref{model}), allows us to proceed beyond the $\omega=0$ limit of
the dilepton rate. As more parameters are determined, the shape of the
soft dilepton spectrum is also better constrained. An interesting open
question is of the number of Fourier coefficients needed to fix the
shape of the dilepton spectrum.  This question is related to the
stability of the transport coefficient, and we plan a study in the near
future to address this question.

The estimates of $\sigma$ from the formula above are subject to lattice
artifacts of order $a^2$ coming from $\vecpf$.  The continuum limit can
then be obtained by an extrapolation in $1/N_t^2$. Finite volume
effects turn out to be invisible within errors. Nor is there any
visible quark mass dependence for small quark masses, since
$m/T_c=0.03$ and $0.1$ give identical results within errors. We
estimate $\sigma/T\approx7\vertex$ in the continuum limit of the
temperature range we studied.  Finally, have used the estimate of
$\sigma$ to predict the soft photon emissivity of the QCD plasma in
equilibrium, as shown in Figure \ref{fg.phot}.

In summary, we adopted a sequence of Bayesian techniques for the inverse
problem of extracting spectral densities, $\rhov$. In view of the
results of \cite{dilepton} we used the dispersion relations for Euclidean
propagators after subtraction of the ideal gas values of the Euclidean
temporal correlator, $\Delta\prop$. We observed that in the QCD plasma, in
the temperature range $1.5\le T/T_c\le3$, the spectral density is peaked
at small energies. This peak was next analyzed using a parametrised form
of the Bayesian prior and the electrical conductivity of the plasma was
extracted. This was then used to predict the soft photon emissivity of
the QCD plasma.

Rough estimates of typical transport related time and length scales in
the QCD plasma can be obtained by using the extracted value of the
electrical conductivity in conjunction with the simple transport
formula $\sigma=C_{EM} n_q\tau_q/m$ ($C_{EM}$ is nothing but the
average charge square: $e^2$). If the number density of quarks, $n_q$,
is substituted by the corresponding entropy density, and the screening
mass used for $m$, then one finds the quark mean free time
$\tau_q\approx0.3$ fm. This is also the time scale for the persistence
of charge, isospin, strangeness and baryon number fluctuations in the
plasma \cite{amy2}. A possible experimental check could be to determine
the mean free path of very soft off-shell photons ($\omega\ll 0.15
(2\pi T)\approx200$ MeV). These are only 20 times longer than $\tau_q$,
\ie, about 6 fm.  The fireball at RHIC may be marginally transparent to
such photons, but at LHC the fireball size could be large enough to
attenuate the intensity of very soft photons.  Such an observation
would constitute direct evidence for short mean free paths in the
plasma.

Some estimates of other transport coefficients can be obtained if one
assumes that the mean free time of gluons is $\tau_q/2$, since they
should be related by colour factors.  Then simple transport formul{\ae}
treated in the same approximation as before lead us to the estimate
that the dimensionless ratio $\eta/S\approx0.2$, where $S$ is the
entropy density of the plasma and $\eta$ is the shear viscosity.  This
ratio is of the same order of magnitude as extracted from present
heavy-ion data \cite{teany} and obeys a bound conjectured in
\cite{son}. It would be useful to make a direct measurement of the
shear viscosity on the lattice.

Many interesting lines of research are relegated to the future.  The
dilepton emissivity away from $\omega=0$ is a conceptually simple
extension, but requires further numerical work, as explained before.
Extending these measurements closer toward $T_c$ where correlation
lengths grow larger \cite{saumen} is of obvious importance, but outside
the scope of this paper, as is the extension to dynamical QCD. The
interesting question of the effectiveness of linear response theory,
and hence of the Kubo formulae closer to $T_c$, can perhaps be probed
using the non-linear susceptibilities defined in \cite{zakopane}.


\begin{thebibliography}{99}
\bibitem{amynote}
   The leading-log approximation gives a negative value of $\sigma$, since
   $g>1$, indicating the failure of the approximation. Dropping the
   logarithm to ``cure'' this problem is unfeasible because the
   expansion is organised in powers of the logarithm.
\bibitem{nullsp}
   The integral on the right of eq.\ (\ref{spec}) could admit a non-trivial
   kernel, \ie, there could be a class of non-vanishing functions for which
   the integral vanishes.  However they must necessarily change sign at
   least once, and we exclude them from the class of admissible spectral
   functions. In this restricted space, eq.\ (\ref{spec}) is invertible
   in the continuum limit. See also \cite{baym}.
\bibitem{integ}
   The discretisation of the integral is made using Newton-Cotes formul\ae
   \cite{press}. The weights due to the conversion of the integral to the sum are
   included in the matrix $K$.
\bibitem{deltafoot}
   The photon production rate in eq.\ (\ref{rate}) contains an extra term in
   $\rho_{00}$ which vanishes but can be shown formally to be of the form
   $\rho_{00}(\omega)= 2\pi\chi_Q\omega\delta(\omega)$, where $\chi_Q$ is
   the charge susceptibility measured in \protect\cite{gavai}. In the rest
   of this paper we treat this rate without this piece. It should be added
   in applications where it is needed.
\bibitem{arvind}
   Considering $\chiv-\chiv^0$ as a function of the $2M$ poles of eq.\
   (\ref{model}) one proves that it is bounded as the poles go to zero
   or infinity in separate groups. The value of this function can then
   be obtained by induction. I would like to thank T.\ Ramadas and Arvind
   Nair for suggesting the method of proof.
\end{thebibliography}
\end{document}